# Cosmic Ray Navigation System (CRoNS) for Autonomous Navigation in GPS-Denied Environments

A. Chilingarian, S. Chilingaryan, and M. Zazyan
A I Alikhanyan National Lab (Yerevan Physics Institute), Alikhanyan Brothers St. 2, Yerevan, Armenia

**ABSTRACT**
In an era where Position, Navigation, and Timing (PNT) systems are integral to our technological infrastructure, the increasing prevalence of severe space weather events and the advent of deliberate disruptions such as GPS jamming and spoofing pose significant risks. These challenges are underscored by recent military operations in Ukraine, highlighting the vulnerability of Global Navigation Satellite Systems (GNSS). In response, we introduce the Cosmic Ray Navigation System (CRoNS). This innovative and resilient alternative utilizes cosmic muon showers for precise location pinpointing, especially in environments where GNSS is compromised or unavailable.
CRoNS capitalizes on an economical, distributed network of compact muon sensors deployed across urban landscapes and potentially integrated into mobile devices. These sensors are tasked with continuously monitoring muon flux resulting from extensive air showers (EASs) triggered by the consistent high-energy particle flux entering Earth's atmosphere. A central AI unit synthesizes the collected data, determining EAS parameters to establish a dynamic reference coordinate system that could span cities and even nations. A notable advantage of CRoNS lies in its capability for reliable operation beneath the Earth's surface and in aquatic environments.
When outfitted with CRoNS, robots, and vehicles gain the ability to autonomously navigate complex urban landscapes and underground passages (≈10 m deep). They achieve this by cross-referencing the measured muon densities against the established reference system, thereby enhancing locational accuracy beyond the potential of singular EAS events.
The adoption of CRoNS promises an effective and scalable backup for PNT services. It leverages the consistent influx of high-energy cosmic protons to create a network of navigational data points. CRoNS paves the way for advancements in navigation technology and the expansion of operational domains beyond the reach of current GPS capabilities.

1.  Introduction

In the intricate weave of modern technology, the Global Navigation Satellite System (GNSS) is indispensable, providing high-accuracy positioning critical for many autonomous and precision applications. GNSS systems' resilience is vital, particularly amidst severe space weather events and other threats that can compromise signal integrity and satellite functionality. The recent exposure of GNSS vulnerabilities to GPS jamming and spoofing, as seen in military operations in Ukraine (Goward, 2023), highlights an urgent need for alternative, resilient navigation solutions. Recent extreme geomagnetic storms indicate an approaching peak in the 25th solar activity cycle, highlighting the need for an autonomous navigation system.
Responding to these vulnerabilities, the US and UK governments initiated programs in 2023 to create sovereign Position, Navigation, and Timing (PNT) frameworks (Department for Science, Innovation and Technology, 2023). These frameworks are poised to support Critical

National Infrastructure (CNI) and serve as reliable backups for key services like defense and emergency responses during GNSS outages.

Various alternative navigation methods are being explored to ensure uninterrupted navigation capabilities, particularly in the face of GNSS compromise due to severe space weather or anti-satellite weaponry. While not yet fully realized, these alternatives offer crucial navigational functions:

- Inertial Navigation Systems (INS): Utilizing accelerometers and gyroscopes, INS calculates position and orientation from speed and movement direction. Common in aircraft, submarines, and spacecraft, they boast immunity to signal jamming due to their independence from external signals.
- Radio Navigation Systems: Precursors to GPS, systems like LORAN and VOR used radio signals to triangulate position. Although many have been decommissioned, they're now being reconsidered as potential GPS backups.
- Celestial Navigation: An age-old art employed by sailors that uses celestial bodies to determine one's position.
- Dead Reckoning is a traditional method for estimating one's current position from a known location by tracking speed, time, and direction.
- Terrestrial Navigation: Relies on landmarks, topographical maps, and compasses, particularly for land navigation.
- Optical Aids: Leverage advanced systems to automate the tracking of celestial bodies for navigation.
- Quantum Navigation: An emerging technology using quantum sensors to provide precise, GPS-independent navigation by measuring acceleration and gravitational fields.

Each method has its advantages and limitations. Combining these systems in military and other critical contexts is common to ensure redundancy and resilience. Quantum navigation using the Bose-Einstein condensate (BEC) of rubidium atoms can be developed as a highly sensitive and precise sensor. The atoms in this state of matter behave coherently, moving in sync, and can be controlled with extreme precision. By manipulating the BEC using magnetic fields and laser beams, researchers can create interference patterns extremely sensitive to external forces, such as changes in gravity or acceleration. These interference patterns can measure the device's rotation, acceleration, and gravitational field in which the condensate is contained. The device can accurately determine its position and orientation in space by comparing the interference patterns generated by the BEC with a reference pattern. This allows for highly precise navigation without the need for traditional GPS signals, making quantum navigation a promising technology for applications in space exploration, autonomous vehicles, and other fields where precise navigation is crucial.

However, achieving the Bose-Einstein condensate state requires cooling atoms to temperatures close to absolute zero (-273°C). The process demands high-precision equipment for cooling, manipulating, and maintaining the condensate, which can be complex to operate and maintain. BECs are highly sensitive to external disturbances such as emerging magnetic fields (of technological or Solar nature), temperature fluctuations, and vibrations. This sensitivity can make maintaining stable conditions for accurate measurements challenging, especially in mobile environments like spacecraft or autonomous vehicles. The condensate

<span style="color:red">must be isolated from external influences, requiring advanced shielding and isolation techniques, which add to the system's complexity and cost. The method relies on analyzing interference patterns, which can be difficult to interpret accurately. Small errors in measurement or environmental changes can lead to significant deviations in results.
The large amount of data generated by the interference patterns requires sophisticated algorithms and processing capabilities to interpret and use the information effectively. The need for precise and stable conditions may limit the equipment's portability, making it less practical for certain applications.</span>

In turn, for navigating and positioning in environments inaccessible to GPS muonography stands out due to its stability, simplicity, and well-known measurements and data analysis methods. This unique technique, operational in subterranean and aquatic settings, utilizes the stable flux of cosmic ray muons for navigational purposes. Its capability was underscored by a Japanese research team deploying a muon-based sensor array in the Tokyo Bay Aqua-Line service tunnel, 45 meters below sea level (Tanaka et al., 2022). In 2022, they successfully used muonography to image a cyclone's vertical profile, revealing variations in density and providing insights that could improve cyclone predictions. Muonography's potential for enhancing autonomous navigation in robots and vehicles, especially in GPS-denied environments, marks it as a significant advancement for resource exploration and geoscientific understanding.

Building on muonography's strengths, the Cosmic Ray Navigation System (CRoNS) is being developed to provide continuous position estimation within a fixed coordinate system. CRoNS utilizes the muonic component of Extensive Air Showers (EASs) generated by high-energy galactic protons with energies between 10 – 1000 TeV. Muons from the primary particles with smaller energies do not reach the ground with amounts allowing reliable recovery of coordinates of shower axes. These muons, capable of penetrating buildings and dispersing up to 100 meters from the shower axes, require hundreds of muon detectors across urban areas. With muon sensors on mobile devices, CRoNS offers a robust framework for accurate distance determination from EAS axes, establishing a new method as a cornerstone of future navigational technologies.

## 2. Cosmic Ray Navigation System (CRoNS)

Utilizing individual muons for precise positioning is challenging due to the extensive random background noise. Archaeological applications, for instance, have required prolonged exposure to large sensors to detect underground cavities (Procureur et al., 2023). We can reduce the random muon background using recovered EAS axes coordinates and achieve high navigation precision.  Although pinpointing the exact location of high-energy particles from a single EAS is not feasible, we can ascertain the radius around the shower axes. These high-energy showers (>0.1 PeV) are frequent and impact areas exceeding 10,000 m². Muons from a single shower, arriving within tens of nanoseconds and typically registered within a millisecond window, enable the recovery shower characteristics. The approximation of the Lateral Distribution Function (LDF), measured by a dense network of muon detectors, facilitates the derivation of five critical shower parameters: coordinates of shower axes on the ground, incidence angles, and the altitude of the primary particle's initial interaction with atmospheric nuclei (Chilingarian et al., 2004a). The accuracy of shower axes determination is influenced by the detector array's density and the primary particle's energy, usually within a 5-10 m range. The continuous and uniform arrival of EASs provides a stream of coordinates with time stamps from an atomic clock, forming a reference coordinate system for mobile vehicle positioning. Leveraging multiple EAS events enhances locational accuracy, mitigating the constraints of singular EAS analysis.

CRoNS offers two modalities for surface and underground positioning. For surface applications, a mobile device's distance from the shower axes is directly deduced by comparing the measured muon density on moving device y with the LDF recovered from the distributed detector network. Subterranean positioning, however, involves estimating the expected muon density after modeling EAS muon propagation through the soil to the target depth. The underground detector's measured muon density is then compared with the estimated LDF. This method does not necessitate a network of underground muon detectors; rather, muon density data from mobile devices can be transmitted to the AI-governed data acquisition system via HF radio, ensuring connectivity up to approximately 10 m depths.

Integrating Neural network models into EAS physics is now dominant, starting from a successful milestone in 2003 (Chilingarian et al., 2004b). A robot or vehicle equipped with a muon detector, comparable to or more advanced than those in the stationary network, can determine its position within the global coordinate system by correlating measured muon densities with those anticipated from registered EASs. While individual EASs fall short of the precision needed for exact location targeting, synthesizing data from the continuous cosmic ray flux substantially enhances positioning accuracy.

CRoNS is underpinned by a distributed network of cost-effective, compact muon sensors strategically positioned atop urban structures or at ground level. The feasibility of integrating CMOS sensors within mobile phones for global EAS monitoring is also being explored (Vandenbroucke et al., 2015). Multiple EASs impact expansive cityscapes so that they can be tracked, establishing a reference system that allows each mobile muon counter to ascertain its precise location. This network of navigational sentinels relentlessly surveys the muon flux from EASs, with the harvested data funneled into a central AI-driven unit that promptly computes EAS parameters. These parameters, encompassing the coordinates of axes, angles of inclination, and the primary particle's energy, merge into an exhaustive and dynamically refreshed reference coordinate system.

### 3. Networks of Muon Detectors on the Earth's Surface

Microfabricated radiation detectors, or integrated circuit radiation sensors, are at the forefront of particle detection technology. These miniaturized devices can identify the presence and characteristics of ionizing radiation. Utilizing advanced micro and nanofabrication techniques akin to those employed in computer chip production, they offer the potential for reduced production costs and seamless integration with other electronic circuits. This integration facilitates functions like synchronization and data transmission directly to the central processing unit (CPU).

Scaling these detectors to form expansive arrays allows for amplifying muon detection capabilities. Constructed from materials such as silicon or germanium, they operate by generating electron-hole pairs in response to ionizing radiation exposure. The quantity of these pairs correlates with the energy of the radiation particle, providing a measure of its intensity. Alternative designs incorporate scintillator materials that emit light upon radiation exposure, which is converted into an electrical signal through a photodetector.

A prime exemplar of this technology is the Timepix chip, a compact, low-power radiation monitoring device equipped with a 300-micron thick silicon sensor boasting a signal threshold of 8 keV/pixel (Granja et al., 2016). The sensor, with its 256 x 256-pixel matrix, covers a total area of 14 x 14 mm² (equivalent to 2 cm²) and can register event count rates up to $10^{11}$ count/(cm² s), rendering it an ideal candidate for distributed muon detection networks. The innovative concept of harnessing smartphones for Extensive Air Shower (EAS) detection hinges on the principle that when high-energy muons impact the sensors within these devices, they effectively transform smartphones into a widespread network of cosmic ray detectors. By leveraging smartphones' precise location and timestamp capabilities, isolating muon counts associated with specific EASs becomes feasible, thereby furnishing valuable data for the real-time recovery of EAS parameters. However, actualizing such a project presents notable challenges, such as mobilizing a substantial participant base and crafting sophisticated software to discern muons from individual EASs.

The CRAYFIS (Cosmic RAYs Found In Smartphones Smartphones, Vandenbroucke et al., 2015) initiative is a groundbreaking endeavor to exploit smartphones' ubiquity to establish an extensive cosmic ray observatory. The initiative seeks to tap into the sensitivity of CMOS sensors in smartphone cameras to high-energy particles, thus enabling the detection of cosmic rays. This approach envisions a global network that could surpass the area coverage of traditional cosmic ray observatories. Furthermore, the project represents a significant contribution to citizen science, empowering individuals to partake in scientific discovery with devices that are part of their everyday lives. Detailed information on commercially available muon detectors and electronics is provided in attachments 1 and 2.

## 4. Simulations of the expected muon flux on the Earth's surface

Muon components of cosmic rays, secondary particles resulting from interactions of primary cosmic rays with the Earth's atmosphere, are detected in significant quantities at the surface. Most of the primary cosmic rays in the energy range 0.1-10 PeV, giving birth to muons, are protons accelerated in supernovae explosions in our galaxy. Figure 1 from (Hovsepyan & Chilingarian, 2023)shows the dependence of primary light nuclei flux (protons and alpha particles) on energy. Notable discrepancies in spectral slope estimates have been observed across different detection methods, including balloon, satellite, and surface array data. These differences can be attributed to the varied experimental techniques, atmospheric effects, a deficit of data at energies above 100 TeV for high-altitude experiments, and inadequate data collection for surface arrays at energies below 500 TeV. Despite these variations, a general concordance can be seen in the energy spectra recorded by the HAWC (Arteaga-Velazqueza, 2021, blue circles) and the high-altitude MAKET-ANI arrays (Chilingarian et al., 2004b, black squares), particularly in the energy range of $5\times10^{13}$ to $10^{15}$ eV.

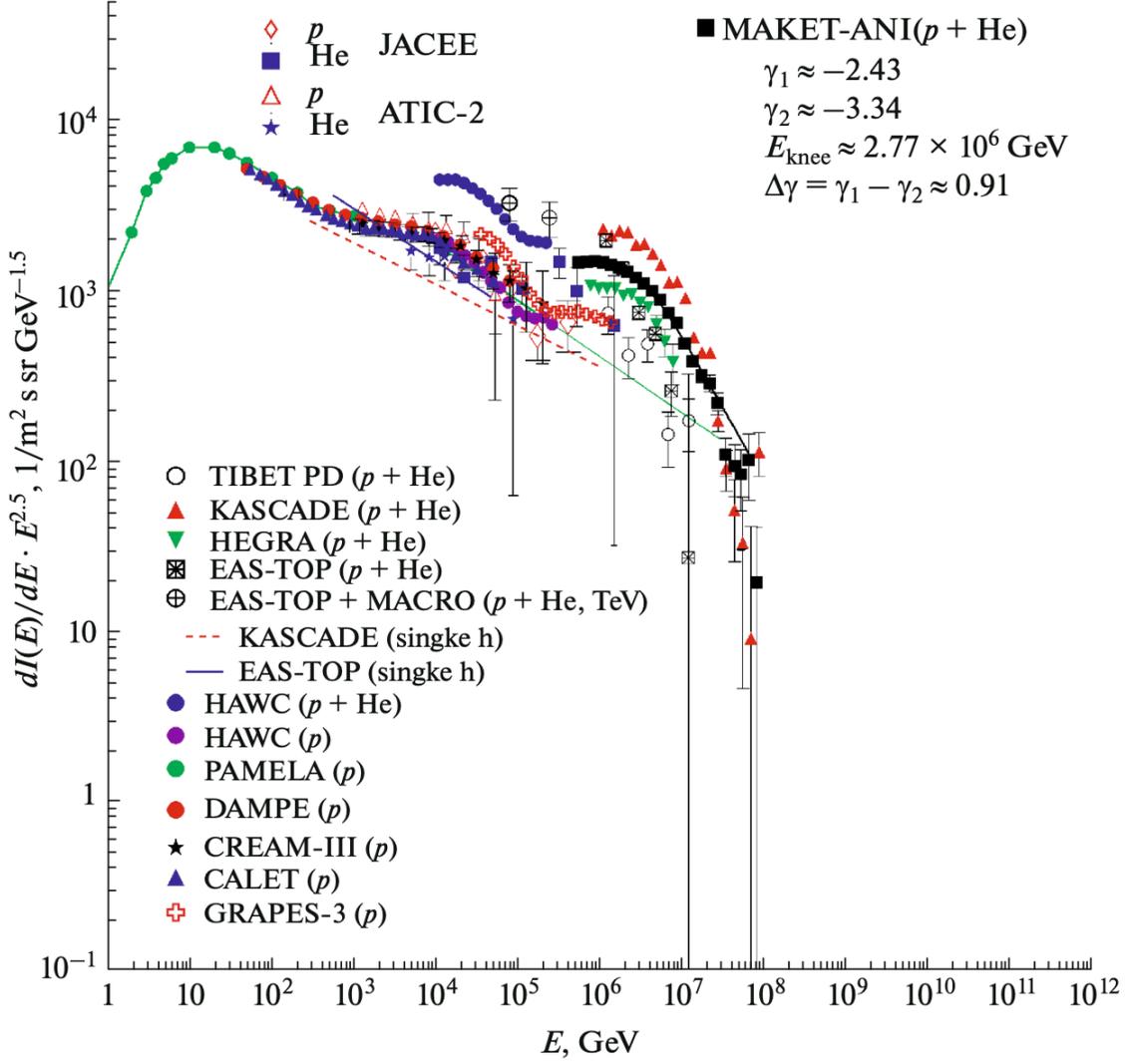

**Figure 1. Primary light nuclei (p+He) spectra measured by the MAKET-ANI (Chilingarian et al., 2004b) detector in comparison with the spectra reported by KASCADE, EAS-TOP, HEGRA, EAS- TOP+MACRO, TIBET experiments (all spectra were obtained with CORSIKA QGSJet01model). The direct balloon measurements by ATIC-2 and JACEE are related to $10^2$ - $10^5$ GeV energies. The modern experiments HAWC (Arteaga-Velazqueza, 2021), PAMELA (Galper et al., 2017), DAMPE (An et al., 2019), CREAM I-III(Yoon et al., 2017), CALET(Adriani et al., 2022), and GRAPES-3 (Varsi et al., 2022) are included as well.**

Typically, the energy spectra of cosmic rays are published in differential form. We aim to assess the expected muon flux at Earth's surface to determine its utility for CRoNS operations. The energy spectrum of primary cosmic rays, specifically protons, in the range of 10 GeV to $10^6$ GeV can be approximated by a power law, $I(E) = A*E^{-\gamma}$, with γ ranging from -2.5 to -2.7 across various energy levels. The normalization constant A, representing the intensity of cosmic ray protons at 1 GeV, is often estimated at 5000 (sec·m²·ster·GeV)⁻¹. We proceed with an assumed spectral index of γ = -2.6. From the differential spectrum, $I(E) = A*E^{-\gamma}$, the integral spectrum is then derived as:

$$J_i(E) = A/(-\gamma+1)*(E_i^{-\gamma+1} - E_{max}^{-\gamma+1}), i=1, N-1.$$

Where N is the number of bins, Emax=$10^7$ GeV is the upper energy limit corresponding to the rare occurrence of galactic protons at the Earth's surface, typically less than once per year per square meter. The approximate equation introduced by Lev Dorman (Dorman, 1975), $J(E) \approx A* E^{-\gamma+1}$ uses the same A and γ values. Table 1 details the expected galactic proton flux over one m², 10,000 m², and one km². The observed frequency of showers at Earth's surface will decrease if we utilize an electron density trigger for the surface array, such as when five particle detectors are activated within a millisecond. Electrons rapidly attenuate in dense atmospheres, so only a small fraction of primary protons with energies up to 100 TeV will generate electron showers that reach the surface. Detectors at altitudes above 4000 m can register such events, although with less than 100% efficiency. However, as part of EASs, muons have a higher probability of reaching the surface in substantial quantities.

**Table 1. The expected number of primary protons hitting the Earth's atmosphere per second at 1m², 10,000 m², and 1 km². In the second column, we post the approximate values obtained with the equation introduced in (Dorman, 1975).**

| Energy (GeV) | Dorman γ = -2.6 | Ntotal(>E) (m$^{-2}$ s$^{-1}$) | Count Rate (10,000 m²) | Count Rate (1 km²) |
|---|---|---|---|---|
| 1 | 62800 | 39270 | 3.9×10$^8$ | 3.9×10$^{10}$ |
| 10 | 1557 | 986 | 9.9×10$^6$ | 9.9×10$^8$ |
| 100 | 39 | 25 | 2.5×10$^5$ | 2.5×10$^7$ |
| 1,000 | 1 | 0.62 | 6.2×10$^3$ | 6.2×10$^5$ |
| 10,000 | 0.025 | 0.016 | 156 | 1.56×10$^4$ |
| 100,000 | 0.0006 | 0.0004 | 3.93 | 393 |
| 1,000,000 | 1.5*10$^{-5}$ | 10$^{-5}$ | 0.1 | 10 |

In Fig. 2, we show the muons reaching the Earth's surface from proton with energy 10 TeV (Fig. 2a) and 1000 TeV (Fig. 2b).

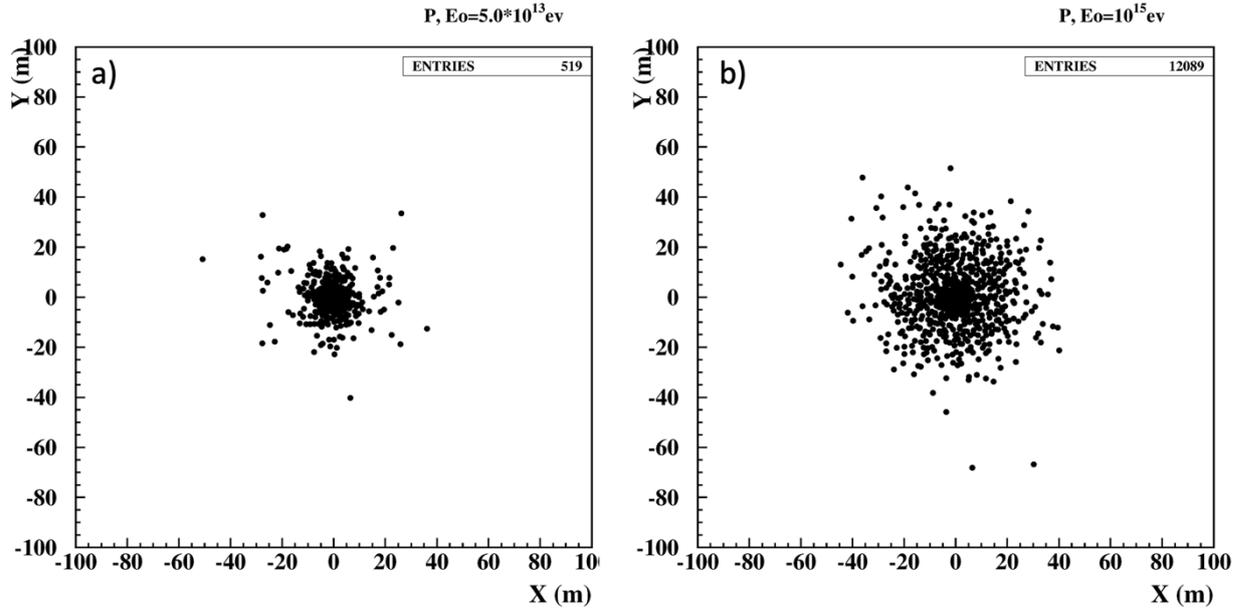

**Figure 2. The distribution of muons hitting the ground from the low-energy primary proton, 10 TeV, a) and 1000 TeV, b).**

To assess the muon rate expected for CRoNS applications, we simulated particle showers over New York City and traced the resulting muons to the surface. Utilizing the CORSIKA simulation code (Heck et al., 1998), incorporating the QGSJETII-04 and UrQMD models for strong interactions. EASs were initiated by primary protons at a direct overhead trajectory (0-degree zenith angle). Muons were tracked until their energy dropped below 10 MeV. Figure 2 illustrates that a significant number of muons reach the ground. With 500 muons impacting an area of 10,000 m² and detected within one millisecond, there is potential for navigation usage. As indicated in Table 1, we expect approximately 156 such muon showers per second, with the number from 1000 TeV protons being considerably fewer (around 0.1) but with a muon count exceeding 10,000. For a larger area of 1 km², the shower frequency increases to 10 per second. Muon showers from intermediate energies will be between these two extreme cases.

Figure 3 shows the lateral distribution of muons at sea level and 15 m. below ground. The spread of muons from the EAS initiated by a 10 TeV primary proton is typically within a 50 m radius. In contrast, the spread from a high-energy proton can extend up to 100 m. For the underground detection, the lateral distribution shrieked ≈ twice. This distribution gives us insight into the necessary detector density for effective CRoNS operations.

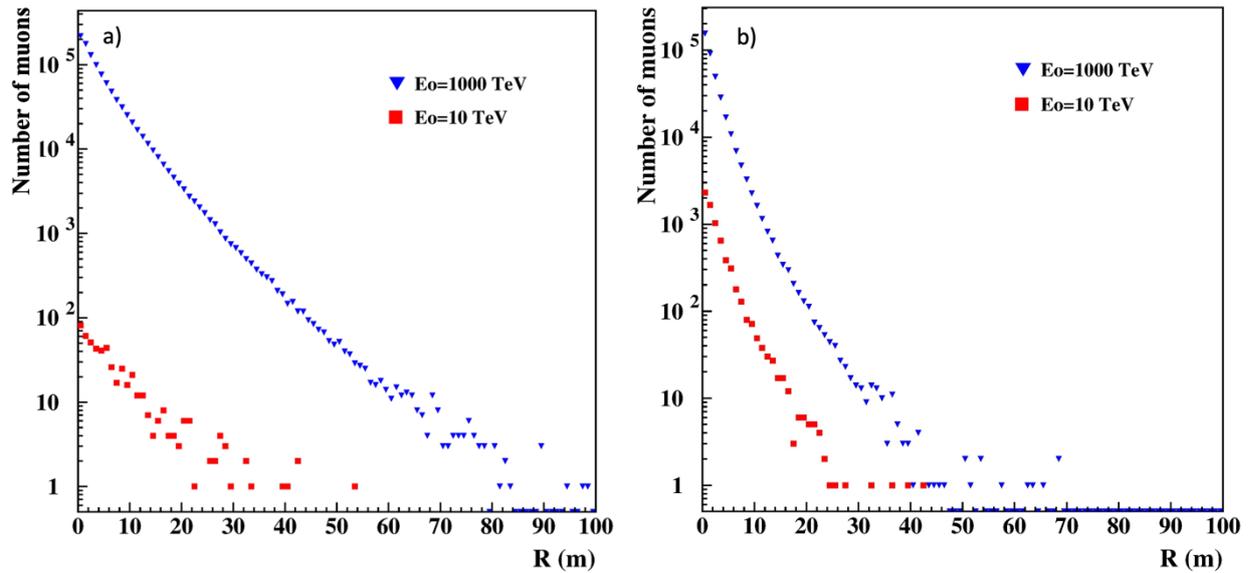

**Figure 3. The lateral distribution of muons with energies above 10 MeV from primary protons with energies 1000 TeV, blue, and 10 TeV, red. One hundred showers for each primary energy were simulated. a) on the Earth's surface; b) under 15 m of soil and concrete (minimal energy of muons 5 GeV).**

It is worth noting that the number of muons of energies less than 10 MeV can be significant, and their spread around the shower axes will be much larger.
In Figure 4, we show the distribution of muons in each EAS for 10 and 1000 TeV energies on the Earth's surface and underground. As expected, the underground muons' distribution is much more compact (E > 5 GeV). The mean number of high-energy muons in a shower is ≈ 10,000, and low-energy muons ≈150. The number of intermediate energy muons will be between these extreme cases.

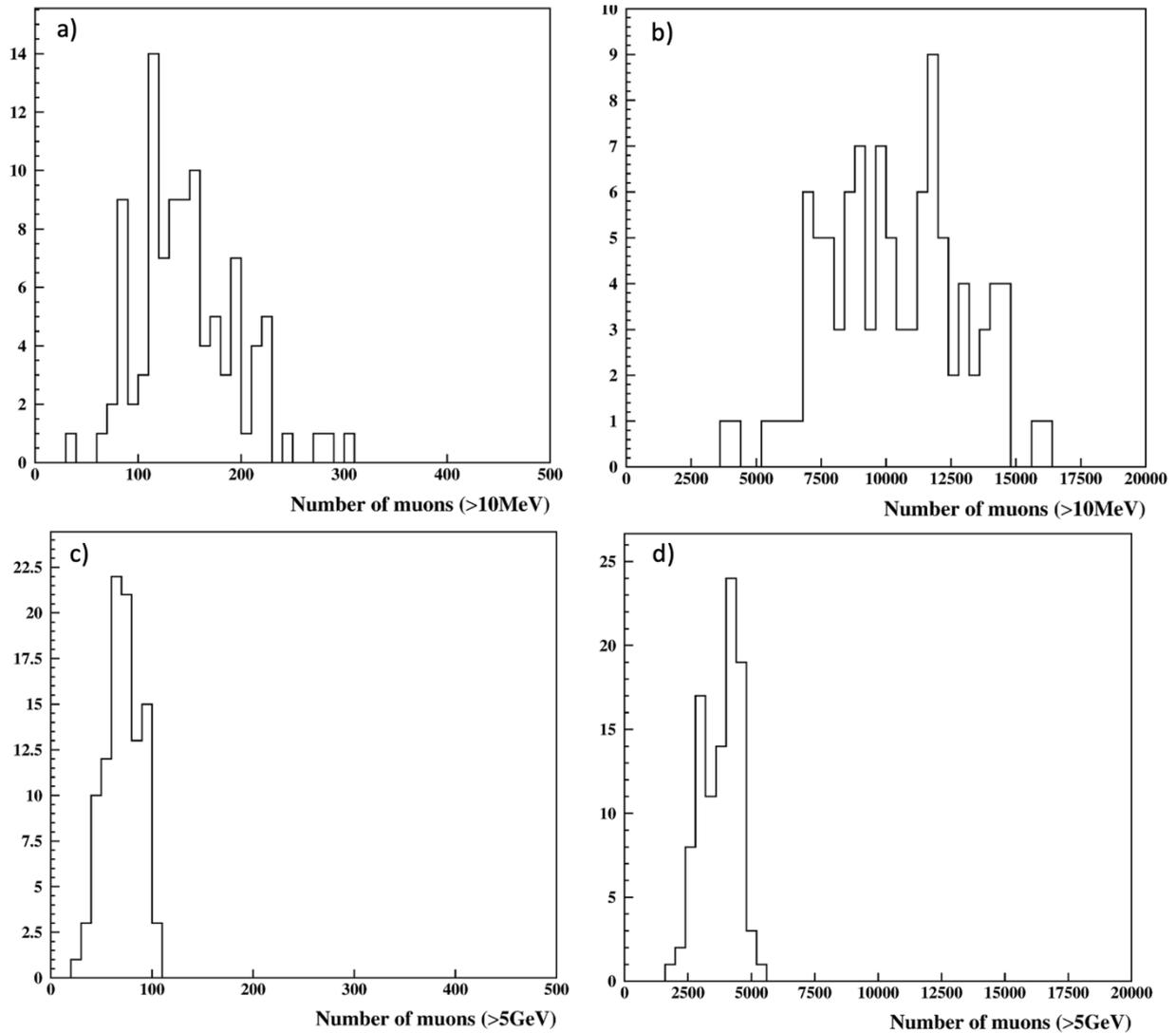

**Figure 4. The distribution of muons from high and low-energy primary protons at sea level (a and b) and underground (c and d).**

5.   **Implementation and Technical Solutions**

The stable flux of galactic protons and nuclei provides enough secondary muons to recover particle type, energy, and shower axes (the center of gravity of muons). Shower parameters will be continuously recovered and stored for further comparison with the muon density measured by detectors on moving vehicles. Monitoring the showers will provide a continuously renewed map of hitting points for vehicle navigation. The scheme of CRoNOS navigation can be looked at as follows:

1. Sensor Deployment and Network Setup

- Locate Muon Sensors: Place muon sensors in areas prepared for surveillance, such as building walls and roofs in urban areas and mobile vehicles.
- Wireless Connection: Install electronics for wireless communication, linking all sensors to a central processor. Connect all detectors within a unified coordinate system.

- Synchronization: Synchronize all sensors with microsecond accuracy.

2. Data Preparation

- Simulation Library: Prepare a library of Extensive Air Shower (EAS) developments in the atmosphere using the CORSIKA code for fixed primary proton energies and angles of incidence. This library will store muon coordinates and energies at specific geographical locations.
- Detector Simulation: Use the GEANT4 package to introduce detector accuracies and all methodological errors, obtaining muon coordinates as measured by the detectors.

3. Neural Network Training

- Simulation Training: Train a neural network using the ANI package (Chilingarian, 1998) or other modern AI packages on collected simulation trials to recover the shower's energy and coordinates based on registered muons.

4. Shower Parameter Recovery

- Shower Coordinate Recovery: Recover the actual shower coordinates and primary proton energy with the trained neural network and measure experimental shower parameters. Store this data within the predefined coordinate system.

5. Mobile Vehicle Navigation

- Muon Measurement: Use the muon measurements from the mobile vehicle to determine its coordinates.
- Reporting: Report the vehicle's coordinates to the driver and operator.

Additional Considerations and Challenges

- Accuracy and Synchronization: Ensuring high accuracy and synchronization among all sensors is crucial for reliable navigation.
- Data Processing: The system requires robust data processing capabilities to handle the real-time data from muon detectors and recover accurate shower parameters.

6. Conclusion

During violent solar flares or upcoming wars, satellite-based systems like GPS might be compromised or unavailable due to anti-satellite weaponry; should be developed navigation methods that can be used. These alternatives can include a Cosmic Ray Navigation System (CRoNS) to create a dynamic local navigation system within big cities and even countries. The distributed system of cheap and compact muon sensors will be located on roofs of tall buildings and continuously measure muon density. EASs containing thousands to millions of muons will trigger the electronics to store particle densities from all sensors. Afterward, AI machinery will quickly estimate the EAS parameters, such as coordinates of shower axes, declination angles, primary particle energy, and others. In the future, robots and vehicles that work independently could become more common in various settings, such as homes, hospitals, factories, and mines. They could also be helpful for search and rescue missions. An additional advantage of the proposed system is the possibility of the autonomous navigation of vehicles and robots in environments where GPS signals are hard to receive. Robots and vehicles equipped with sophisticated muon detectors navigate the concrete jungles and the unseen terrains below the Earth's surface by cross-referencing measured muon densities against the established reference system. In an age where the fabric of global navigation is under threat from the specter of technological warfare, a revolutionary navigation method immune to the vulnerabilities of satellite-based systems is not just a boon but a necessity. The Cosmic Ray Navigation System (CRoNS) emerges as a vanguard of this new era, offering a dynamic and robust navigation system crafted to withstand the challenges of modern combat environments and beyond. Dear Team,
Please take note of the following information:

<span style="color:red">Attachment 3 provides the initial cost estimate for the 1 km² CRoNOS project. The project is scalable. For a 10 km² CRoNOS project (large city), an additional 4000 muon detectors need to be installed. The extra costs for software tuning, simulations, and network training will be minimal.</span>

**Attachment 1. Examples of available muon detectors**

1. CosmicWatch Desktop Muon Detector (Dujmic et al., 2017)

- Description: This simple, low-cost muon detector is designed for educational purposes but can be adapted for other uses.
- Features:
    o Compact design.
    o Uses a scintillator and silicon photomultiplier (SiPM) to detect muons.
    o Provides real-time data on muon counts.
    o USB interface for data transfer.
- Cost: Approximately $100-$200 per unit.

2. LND 712 Geiger-Müller Tube (LND, 2024)

- Description: A Geiger-Müller tube can be used as the core of a muon detector. The LND 712 is a popular choice for detecting cosmic rays.
- Features:
    o Sensitive to beta and gamma radiation, including muons.
    o Compact and durable.
    o It can be integrated with custom electronics for muon detection.
- Cost: Approximately $80-$120 per tube.

3. MuonPi (Stradolini et al., 2017)

- Description: A Raspberry Pi-based muon detector designed for educational and research purposes.
- Features:
    - Uses a scintillator and SiPM for muon detection.
    - Connects to a Raspberry Pi for data processing and logging.
    - Compact and easy to deploy.
- Cost with computer annd electronics: Approximately $200-$300 per unit.

Example Setup

1. CosmicWatch Desktop Muon Detector (Dujmic et al., 2017)
2. Microcontroller: Use an ESP32 for local data processing and wireless data transfer.
3. Data Aggregation: Use a central server or Raspberry Pi to collect and analyze data from multiple detectors.

**References**

Dujmic, B., Gilbert, J., & Tiffenberg, J. (2017). CosmicWatch: The low-cost muon detector you can build yourself. *Physics Education*. Retrieved from https://iopscience.iop.org/article/10.1088/1361-6552/aa9e9e

LND Inc. product catalog and specifications.

Stradolini, F.,Tuoheti, A., Paolo Motto Ros, P.M., et al. (2017) Conference: New Generation of CAS (NGCAS).  10.1109/NGCAS.2017.67

**Attachment 2. Electronics solutions**

Wireless Data Transfer

1. Wi-Fi 6 (802.11ax) Modules:
    - Intel Wi-Fi 6 AX200: Provides high-speed wireless data transfer with low latency.
    - Qualcomm QCA6390: A versatile module supporting Wi-Fi 6 and Bluetooth 5.1.
2. 5G NR (New Radio) Modules:
    - Qualcomm Snapdragon X55 5G Modem: Offers high-speed data transfer with low latency, suitable for real-time applications.
    - Huawei Balong 5000 5G Modem: Another fast and reliable 5G connectivity option.
3. LoRaWAN (Long Range Wide Area Network):
    - Semtech SX1276: Suitable for long-range, low-power applications where high data transfer speed is not the primary requirement, but coverage is crucial.
4. Zigbee / Thread for IoT:
    - TI CC2652: Ideal for low-power, short-range communication between sensors and central nodes.

Synchronization

1. GPS Disciplined Oscillators (GPSDO):
   - Jackson Labs Fury GPSDO: Provides precise timing synchronization with microsecond accuracy using GPS signals.
   - Trimble Thunderbolt E GPSDO: Another reliable GPSDO for accurate timing.
2. IEEE 1588 Precision Time Protocol (PTP) Devices:
   - Microsemi TimeProvider 2700: Supports IEEE 1588 PTP for precise network time synchronization.
   - Meinberg LANTIME M400: A PTP grandmaster clock for accurate time distribution.
3. White Rabbit (WR) Timing System:
   - Seven Solutions White Rabbit Switch: Combines Ethernet with sub-nanosecond synchronization, suitable for high-precision applications.

Additional Components

1. Microcontrollers and Processors:
   - ESP32: A powerful microcontroller with integrated Wi-Fi and Bluetooth, suitable for wireless communication and synchronization.
   - Raspberry Pi: A versatile single-board computer that can be used for data processing and managing wireless communications.
2. Real-Time Clocks (RTC):
   - DS3231: A high-precision RTC module with a temperature-compensated crystal oscillator (TCXO).
3. FPGA for Custom Timing Solutions:
   - Xilinx Zynq-7000: Allows for creating high-precision custom timing and synchronization solutions.

Implementation specifications

- Network Topology: Implement a hybrid network topology combining Wi-Fi 6 or 5G for high-speed data transfer with LoRaWAN for extended coverage.
- Synchronization Backbone: Use GPSDO or PTP devices to establish a precise time synchronization backbone across all sensors and central nodes.
- Integration with AI: Ensure that your data processing units (like Raspberry Pi or FPGA) are integrated with the neural network for real-time data analysis and decision-making.

Example Setup

1. Sensor Nodes: Equipped with ESP32 for local data processing and wireless communication and DS3231 for local timing.
2. Central Processor: A server or a powerful single-board computer (like Raspberry Pi) connected to a White Rabbit Switch for network-wide synchronization and data aggregation.
3. Data Transmission: To ensure high-speed data transfer, use 5G modems for mobile units and Wi-Fi 6 for stationary units.
4. Timing Synchronization: Deploy GPSDO at fixed points and PTP devices to ensure synchronized operations across the network.

**Attachment 3. Preliminary Budget Table for CRoNOS Project (18 Months)**

| Item | Quantity | Unit Cost (USD) | Total Cost (USD) |
|---|---|---|---|
| **Muon Detectors** | | | |
| CosmicWatch Muon Detector | 400 + 40 spares | $200 | $88,000 |
| **Additional Electronics** | | | |
| Wireless Module (ESP32) | 400 + 40 spares | $10 | $4,400 |
| Power Supply (Battery/Solar) | 400 + 40 spares | $20 | $8,800 |
| Synchronization Module (GPS) | 400 + 40 spares | $15 | $6,600 |
| **Cabling and Infrastructure** | | | |
| Cabling (Ethernet, Power) | - | - | $15,000 |
| **Installation** | | | |
| Installation Cost | 400 | $50 | $20,000 |
| **Central Processing Unit** | | | |
| Robust Central Processor | 1 | $50,000 | $50,000 |
| Spare and Backup System | 1 | $50,000 | $50,000 |
| **Software Development** | | | |
| Software Development Team (18 months) | 1 | $260,000 | $260,000 |
| Simulation Software Integration | 1 | $50,000 | $50,000 |
| Neural Network Training | 1 | $40,000 | $40,000 |
| **Simulation and Data Preparation** | | | |
| Computational Resources | 1 | $20,000 | $20,000 |
| Computational Scientist (18 months) | 1 | $120,000 | $120,000 |
| **Data Storage and Management** | | | |
| Cloud Storage (10 TB/year) | 30 TB | $1,000 | $30,000 |
| Database Management System Setup | 1 | $10,000 | $10,000 |
| Data Backup and Redundancy | 1 | $5,000 | $5,000 |
| Database Administrator (18 months) | 1 | $105,000 | $105,000 |
| **Project Management and Support** | | | |
| Project Manager (18 months) | 1 | $180,000 | $180,000 |
| Secretary (18 months) | 1 | $90,000 | $90,000 |
| Technical Assistants (2, 18 months) | 2 | $120,000 each | $240,000 |
| **Business Travel and Logistics** | | | |
| Travel Expenses | - | $45,000 | $45,000 |
| Logistics (Shipping and Handling) | - | $30,000 | $30,000 |
| **Maintenance and Ongoing Costs** | | | |

| Item | Quantity | Unit Cost (USD) | Total Cost (USD) |
|---|---|---|---|
| Annual Software Maintenance | 1.5 years | $30,000 | $30,000 |
| Annual Data Storage | 30 TB | $30,000 | $30,000 |
| Part-time Developer/Engineer | 1.5 years | $75,000 | $75,000 |

Total Initial Costs

| Category | Total Cost (USD) |
|---|---|
| Muon Detectors | $88,000 |
| Additional Electronics | $19,800 |
| Cabling and Infrastructure | $15,000 |
| Installation | $20,000 |
| Central Processing Unit | $100,000 |
| Software Development | $350,000 |
| Simulation and Data Preparation | $160,000 |
| Data Storage and Management | $180,000 |
| Project Management and Support | $510,000 |
| Business Travel and Logistics | $75,000 |
| Maintenance and Ongoing Costs | $135,000 |
| **Total Initial Cost** | **$1,652,800** |

Total Annual Ongoing Costs

| Category | Total Cost (USD) |
|---|---|
| Annual Software Maintenance | $20,000/year |
| Annual Data Storage | $10,000/year |
| Part-time Developer/Engineer | $50,000/year |
| **Total Annual Ongoing Cost** | **$80,000/year** |